\documentclass[%
 preprint, 
 amsmath,amssymb,
 aps, physrev,
floatfix,
]{revtex4-2}
\usepackage{xcolor}
\usepackage{graphicx}
\usepackage{dcolumn}
\usepackage{bm}

\begin{document}

\preprint{APS/123-QED}

\title{\textbf{Four-channel prototype using coherent combining of ultrashort laser pulses for dipole configuration approximation.} 
}%

\author{Konstantin Burdonov}
\email{Contact author: k.burdonov@ipfran.ru}
\affiliation{Federal Research Center A.V. Gaponov-Grekhov Institute of Applied Physics of the Russian Academy of Sciences, Nizhny Novgorod, Russia}

\author{Mikhail Zolotavin}
\affiliation{Federal Research Center A.V. Gaponov-Grekhov Institute of Applied Physics of the Russian Academy of Sciences, Nizhny Novgorod, Russia}

\author{Alexey Sidnev}
\affiliation{Federal Research Center A.V. Gaponov-Grekhov Institute of Applied Physics of the Russian Academy of Sciences, Nizhny Novgorod, Russia}

\author{Evgeny Blinov}
\affiliation{Federal Research Center A.V. Gaponov-Grekhov Institute of Applied Physics of the Russian Academy of Sciences, Nizhny Novgorod, Russia}

\author{Alexander Kotov}
\affiliation{Federal Research Center A.V. Gaponov-Grekhov Institute of Applied Physics of the Russian Academy of Sciences, Nizhny Novgorod, Russia}

\author{Sergey Perevalov}
\affiliation{Federal Research Center A.V. Gaponov-Grekhov Institute of Applied Physics of the Russian Academy of Sciences, Nizhny Novgorod, Russia}

\author{Artem Korzhimanov}
\affiliation{Federal Research Center A.V. Gaponov-Grekhov Institute of Applied Physics of the Russian Academy of Sciences, Nizhny Novgorod, Russia}

\author{Ivan Mukhin}
\affiliation{Federal Research Center A.V. Gaponov-Grekhov Institute of Applied Physics of the Russian Academy of Sciences, Nizhny Novgorod, Russia}

\author{Alexey Pestov}
\affiliation{The Institute for Physics of Microstructures of the Russian Academy of Sciences, Nizhny Novgorod, Russia}

\author{Mikhail Starodubtsev}
\affiliation{Federal Research Center A.V. Gaponov-Grekhov Institute of Applied Physics of the Russian Academy of Sciences, Nizhny Novgorod, Russia}

\author{Efim Khazanov}
\affiliation{Federal Research Center A.V. Gaponov-Grekhov Institute of Applied Physics of the Russian Academy of Sciences, Nizhny Novgorod, Russia}

\author{Andrey Shaykin}
\affiliation{Federal Research Center A.V. Gaponov-Grekhov Institute of Applied Physics of the Russian Academy of Sciences, Nizhny Novgorod, Russia}

\author{Alexander Soloviev}
\affiliation{Federal Research Center A.V. Gaponov-Grekhov Institute of Applied Physics of the Russian Academy of Sciences, Nizhny Novgorod, Russia}

\date{\today}

\begin{abstract}
This paper presents a four-channel prototype system for the geometric combining and coherent addition of tightly focused femtosecond laser radiation into a standing-wave field configuration. A stabilization system for beam pointing and relative phase of the four optical channels has been implemented, and its performance has been experimentally demonstrated. To characterize the standing-wave electromagnetic field distribution at the main focus of the system, an original measurement technique based on a fiber subwavelength optical probe has been employed. This work has been conducted in support of the exawatt-scale XCELS project.
\end{abstract}

\maketitle



The XCELS project (eXawatt Center for Extreme Light Studies) \cite{Mukhin_2021,Khazanov2023} aims to construct a unique multi-channel laser system capable of reaching exawatt-level peak power, enabling the generation of QED particle cascades from the physical vacuum \cite{PhysRevLett.101.200403,PhysRevLett.105.080402,PhysRevSTAB.14.054401}. At the output of each of the twelve channels, featuring a square aperture exceeding 600 mm, femtosecond, optically phase-locked laser pulses with a peak power greater than 50 PW are expected. The prototype for an individual channel is the PEARL petawatt laser facility \cite{Ginzburg:21,Soloviev:2024}, developed at the Institute of Applied Physics of the Russian Academy of Sciences (Nizhny Novgorod).

\begin{figure}
    \centering
    \includegraphics[width=0.5\textwidth]{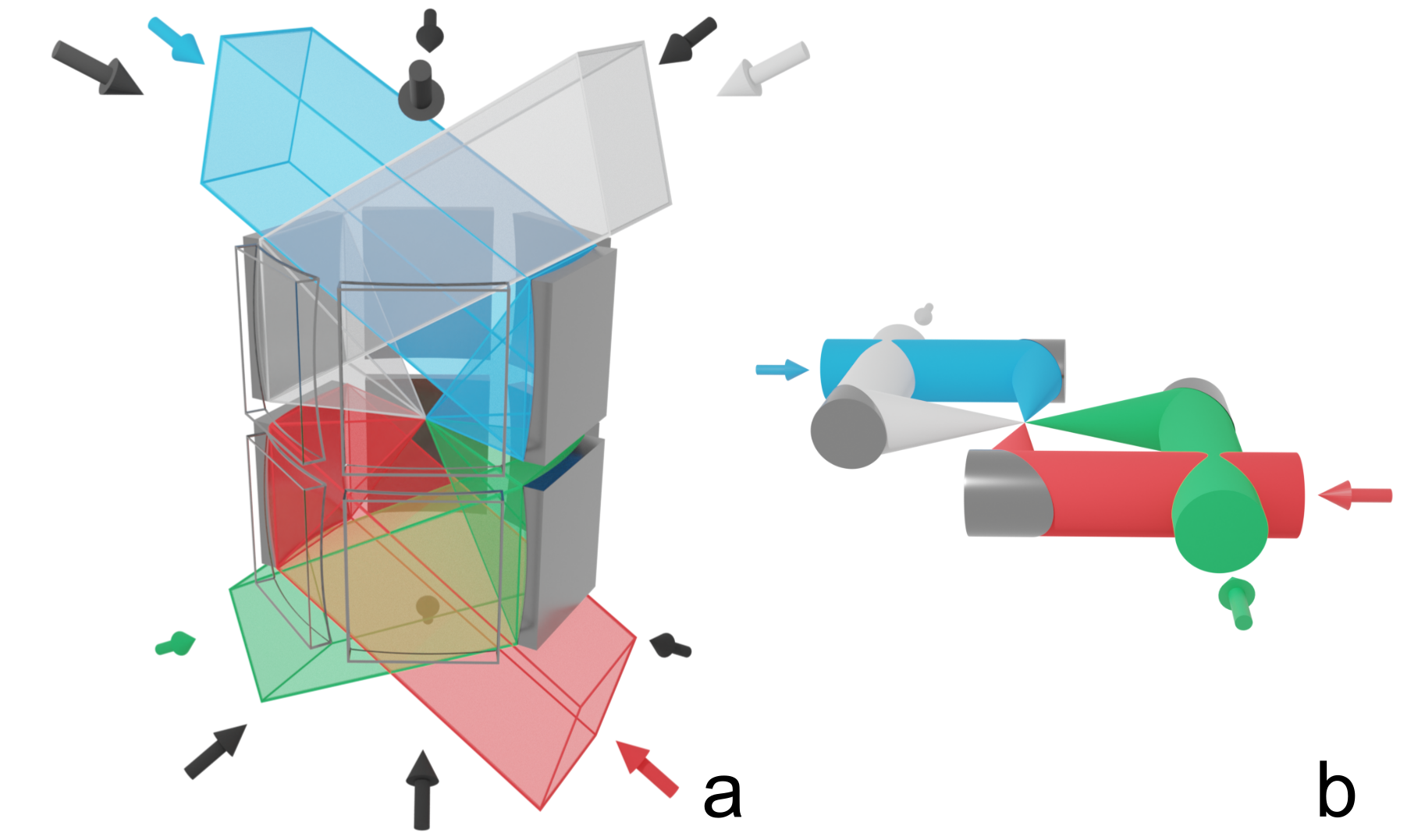}
    \caption{Schematic geometry of the XCELS twelve-channel facility (a); geometry of the four-channel prototype, topologically equivalent to four of the twelve channels of the XCELS set-up (b).}
    \label{fig:XCELS}
\end{figure}

A key feature of the XCELS project is the implementation of the dipole focusing concept \cite{PhysRevA.86.053836}. Using two tiers of $\sim f/1$ parabolic mirrors \cite{Khazanov2023}, the beam from each channel is focused into the main focus of the system, as illustrated in Fig.~\ref{fig:XCELS}(a). Upon coherent combining, this arrangement generates a field topology approaching that of a dipole wave \cite{doi:10.1080/713821943,PhysRevLett.104.220404,PhysRevA.86.053836,PhysRevLett.111.060404}, which yields the theoretical maximum optical field strength for a given laser power \cite{Sidnev}.

Numerical simulations \cite{Efimenko2018,PhysRevX.7.041003,Bashinov_2018,Efimenko2023_1,Efimenko2023_2,PhysRevE.99.031201} indicate that a laser field in an electric dipole configuration can trigger QED particle cascades at a total peak power of slightly below 10~PW. The low cascade generation threshold in a dipole wave is attributed not only to the maximization of the electric field at the focus, but also to the standing-wave nature of the field. The latter facilitates prolonged confinement of seed electrons in the focal region, enhances the radiation reaction power, and consequently increases the number of gamma-ray photons participating in the Breit--Wheeler process.

Coherent combining of multiple tightly focused, extreme-power laser channels into a dipole field configuration is a highly ambitious endeavor that requires the integrated solution of several scientific and engineering challenges: wavefront correction for individual channels, fine active stabilization of beam pointing and relative phases, and precise focusing from multiple directions onto a single point. It should be noted that methods and approaches for correcting the spatial phase of high-power single-channel femtosecond laser systems have been extensively discussed in previous studies \cite{Soloviev_2020,Kotov_2021,soloviev2022improving,kotov2024improving}.

Experimental demonstrations of coherent combining across a large number of channels have been reported previously, primarily for moderate-power fiber-optic systems \cite{Mu:16,Peng:17,Klenke:18,Li:18,Liu:18,Peng_2018,Zhou:18,Wang:19,Peng:19,Chang:20,Fsaifes:20,Wang2021,9635681}. However, these experiments typically employed paraxial scalar focusing to generate a traveling-wave field configuration, where the feedback signal can be readily acquired using standard focal-plane arrays, such as CCD/CMOS cameras. Such detectors, however, are unsuitable for standing-wave field configurations, as they cannot capture radiation arriving from counter-propagating directions.

\begin{figure}
    \centering
    \includegraphics[width=0.5\textwidth]{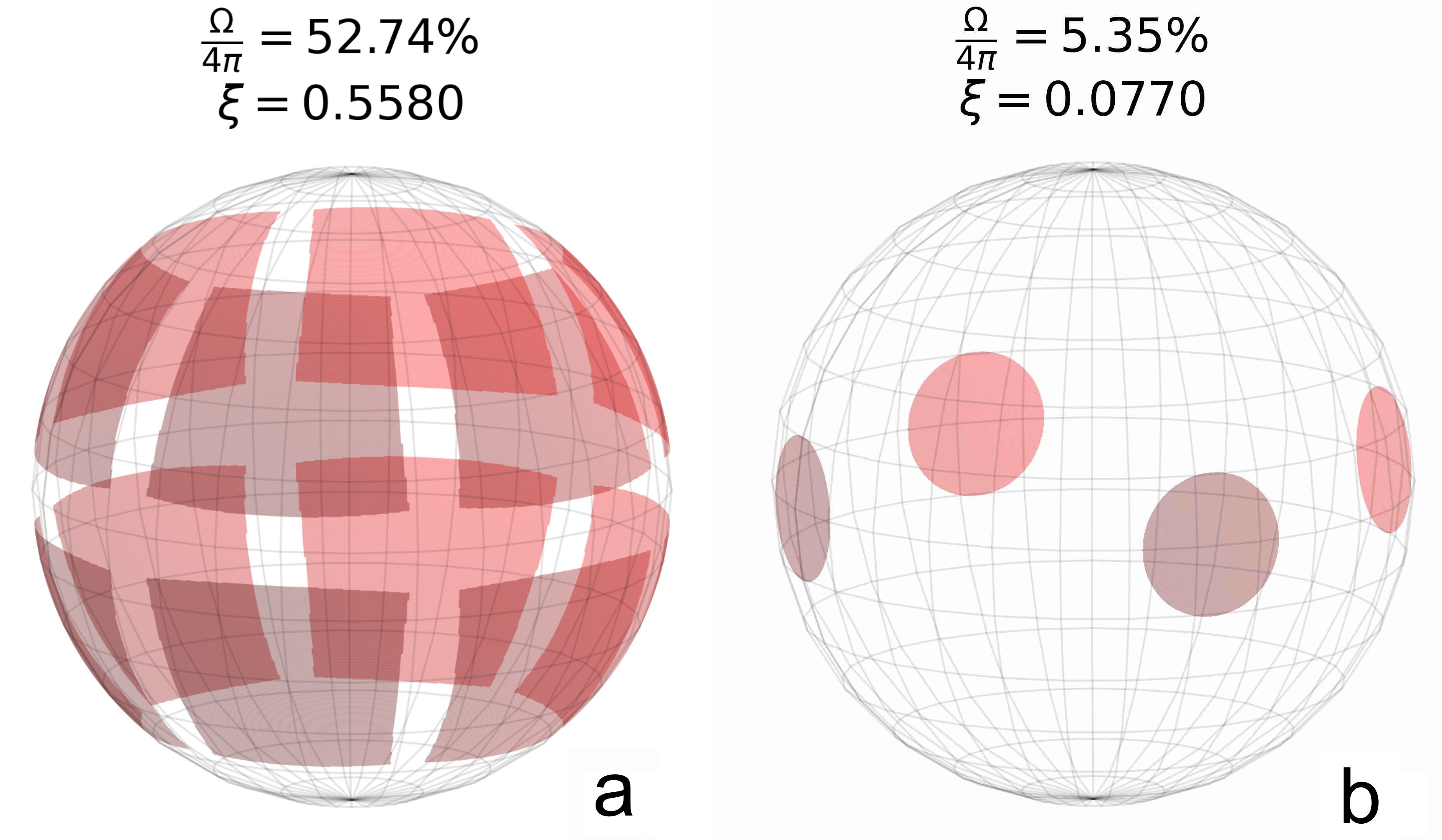}
    \caption{Celestial sphere coverage by converging beams in the experimental approximation of a dipole wave for: (a)  the 12-beam topology of the XCELS project and (b) the prototype (4 circular beams with $f/2$ focusing). Above each panel, the corresponding values of the parameter $\xi = E^2/E_d^2$~\cite{Sidnev} - which quantifies the proximity of the squared electric field $E$ to the theoretical limit for an electric dipole configuration, $E_d = (8\pi/3)\sqrt{P}/\lambda$~\cite{doi:10.1080/713821943,PhysRevA.86.053836}---and the fractional celestial sphere coverage $\Omega/(4\pi)$ by the converging radiation are indicated.}
    \label{fig:Sph_1}
\end{figure}

In this article, we present a four-channel prototype of the XCELS coherent combining system, which provides an approximation of a standing dipole wave. The prototype topology (see Fig.~\ref{fig:XCELS}(b)), first introduced in Ref.~\cite{Burdonov2024}, corresponds to four of the twelve XCELS channels arranged in a single plane (see Fig.~\ref{fig:XCELS}(a)). This configuration enables the testing of all key optical components, as well as the stabilization principles and coherent combining scheme. As expected,  compared to the target twelve-beam geometry (Fig.~\ref{fig:Sph_1}(a)), the prototype yields a less accurate approximation of the dipole wave due to the reduced number of channels and their sparser packing on the celestial sphere (see Fig.~\ref{fig:Sph_1}(b)). Furthermore, for simplicity, the prototype omits the stretcher - compressor system and amplifier stages inherent to the CPA (chirped-pulse amplification \cite{STRICKLAND1985219}) technology. 

Our work focuses on the multi-beam approximation of a dipole wave - a standing-wave field configuration characterized by vectorial features and a fundamental difficulty in generating a feedback signal for channel combining using conventional methods. Measurements of the interference field at the focus were performed using a custom-developed subwavelength probe. Consequently, among other contributions, we introduce a novel method for measuring optical field interference structures, specifically tailored for standing-wave configurations arising from tight focusing and/or multi-beam focusing with a high numerical aperture.



\begin{figure}
    \centering
    \includegraphics[width=0.85\textwidth]{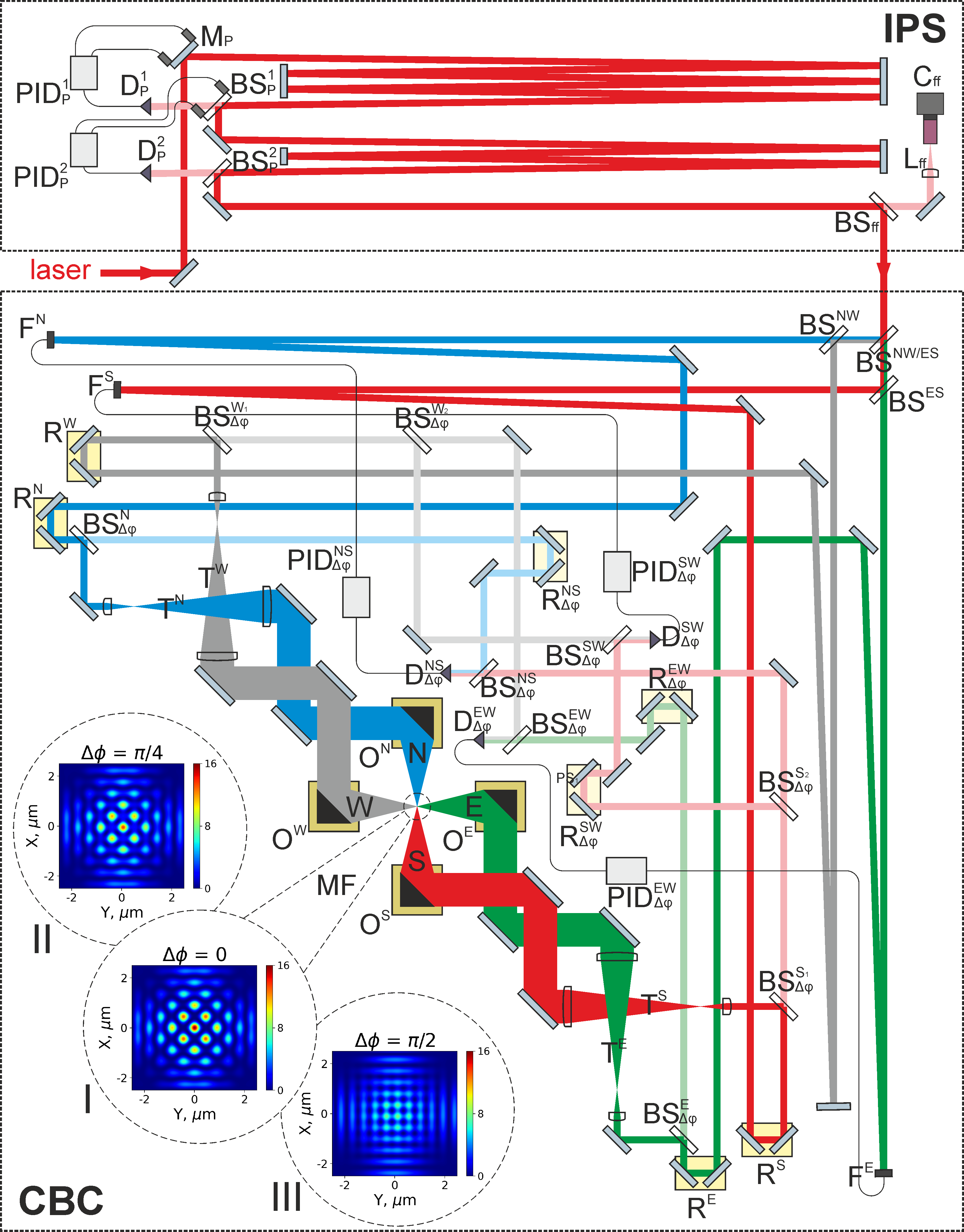}
    \caption{Experimental schematic of the four-beam combining prototype system. Key functional modules are highlighted by rectangles: IPS---Input beam Pointing Stabilization; CBC---Coherent Beam Combining. Inset circles show the calculated interference structures: (I) for zero relative phase between N-S and E-W pairs -- 4E-configuration, (II) for a phase shift of $\pi$/4, and (III) for a phase shift of $\pi$/2 -- 2$\sqrt2$E-configuration.}
    \label{fig:set_up}
\end{figure}

Figure~\ref{fig:set_up} shows the optical schematic of the prototype. The source was a mode-locked Ti:sapphire femtosecond laser operating at a wavelength of 910~nm, with a pulse duration of 30~fs (full width at half maximum, FWHM) and a repetition rate of 70~MHz, delivering an average power of approximately 0.4~W. The transverse intensity profile of the input beam was quasi-uniform, with a circular aperture of approximately 1~cm in diameter.
The laser average power stability was better than 3\% RMS (root mean square) over the measurement durations.

The setup comprised two modules, outlined in the figure with black dashed rectangles: IPS (Input beam Pointing Stabilization) and CBC (Coherent Beam Combining). The IPS module was constructed using a mirror M$_{P}$ and a beam splitter BS$^1_{P}$ arranged in series, both integrated into a feedback loop utilizing PID controllers. 

Within the CBC module the beam was divided into four channels: N (blue), S (red), E (green), and W (gray), according to the directions from which the beams are directed into the main focus (MF). The separation was carried out using three beam splitters: BS$^{NW/ES}$, BS$^{NW}$, and BS$^{ES}$. Here and further in the text, the superscripts N, S, E and W indicate that the element belongs to the corresponding channel.

Each channel incorporated the following components: a mechanical delay line, R (raw delay line), for coarse optical path length matching; a piezo-mounted mirror, F (fine delay line), for fine-tuning the optical path lengths to achieve phase stabilization; 
and a 5$\times$ beam-expanding telescope, T, followed by an off-axis parabolic mirror, O, which provides $f/2$ focusing into the MF. Designations for high-reflectivity transport mirrors are omitted for clarity.

The four channels were directed to the MF in a pairwise counter-propagating configuration in such a way that the counter-propagating pairs of N-S and E-W are perpendicular to each other (see Inset of Figure 3 \ref{fig:set_up}). The interference field in the vicinity of the MF was characterized using a subwavelength optical probe. 

The aim for the four-beam prototype is an interference structure with the maximum squared electric field, corresponding to codirectional 
polarization and zero relative phase of all four channels at the MF point (Fig. \ref{fig:set_up} (I)). From here on, we will call this situation the 4E-configuration.
The minimum value of the maximum field amplitude is achieved when the relative phase of the standing waves formed by pairs of opposing channels is shifted by $\pi$/2 (Fig. \ref{fig:set_up} (III)). This situation will be conventionally referred to as the 2$\sqrt2$E-configuration. 2$\sqrt2$E also occurs when the polarizations of the channel pairs forming the standing waves are orthogonal to each other.
In the experiment, the transition between 4E and 2$\sqrt2$E is accomplished by changing the setpoint voltages on the piezoelectric actuators F.



At the entrance of the CBC module the beam was stabilized by the IPS system (see Fig.~\ref{fig:set_up}) prior to being split into four independent channels. The IPS system simultaneously stabilized the transverse positions of the beam spots at the surfaces of quadrature photodiodes D$^1_P$ and D$^1_P$, located at a significant distance from each other. This arrangement enabled stabilization of both the beam position and its direction. Stabilization was achieved by mechanically tip-tilting the mirror M$_{P}$ and beam splitter BS$^1_P$. The angular piezo actuators were driven using PID$^1_{P}$ and PID$^2_{P}$ controllers that processed the photodiode signals in real time.

\begin{figure}
    \centering
    \includegraphics[width=0.85\textwidth]{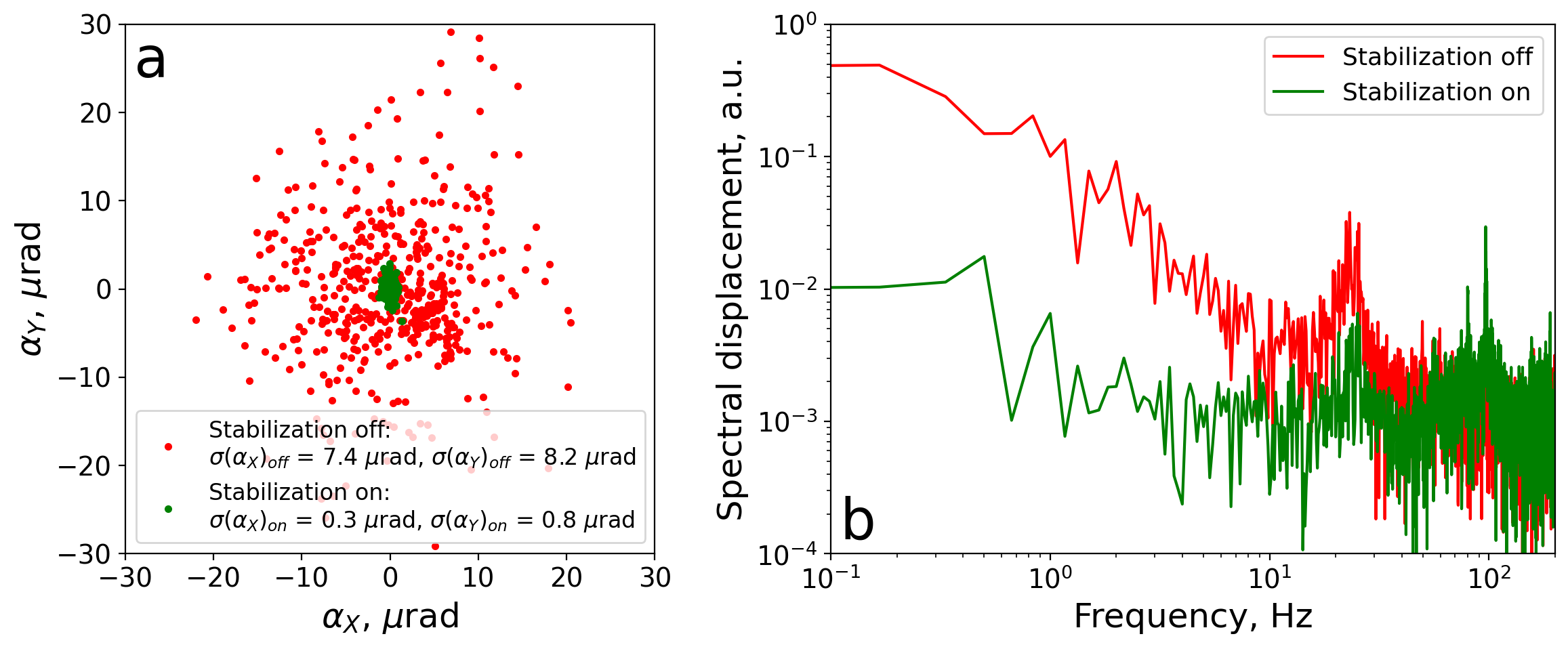}
    \caption{Angular displacement of the beam recorded by camera C$_{ff}$ (a) and the frequency spectrum of the beam displacement along the $X$-axis (b).}
    \label{fig:stability_NF_FF}
\end{figure}

The performance of the IPS is illustrated in Figure \ref{fig:stability_NF_FF} (a), which shows the far-field beam displacements at the CBC input, measured at 5 Hz, for the IPS system on (green dots) and off (red dots). The measurements were made with camera C$_{ff}$, located in the focal plane of lens L$_{ff}$. In our specific design, active stabilization reduced the beam angular displacements by an order of magnitude: from 8~\textmu rad to 0.3~\textmu rad horizontally, and from 7.4~\textmu rad to 0.8~\textmu rad vertically.
The frequency spectrum of the far-field displacements 
is shown in Figure \ref{fig:stability_NF_FF} (b). It can be seen that the system effectively suppresses beam position displacements up to frequencies of approximately 100 Hz.

The XCELS project itself \cite{Khazanov2023} plans to equip each of the combined optical channels with a stabilization system similar to the IPS system described in this work.


The CBC included three relative phase stabilization subsystems for channel pairs based on mirrors F (see Fig. \ref{fig:set_up}), mounted on piezoelectric positioners driven by the controllers PID$_{\Delta \phi}$. The feedback signal was derived from the interference of leakage beams detected by photodiodes D$_{\Delta \phi}$. These leakage beams are tapped off after beam splitters BS$_{\Delta \phi}$ and are depicted in a lighter shade in the schematic.
Channel W served as the reference; its phase was not actively controlled.

\begin{figure}
    \centering
    \includegraphics[width=0.45\textwidth]{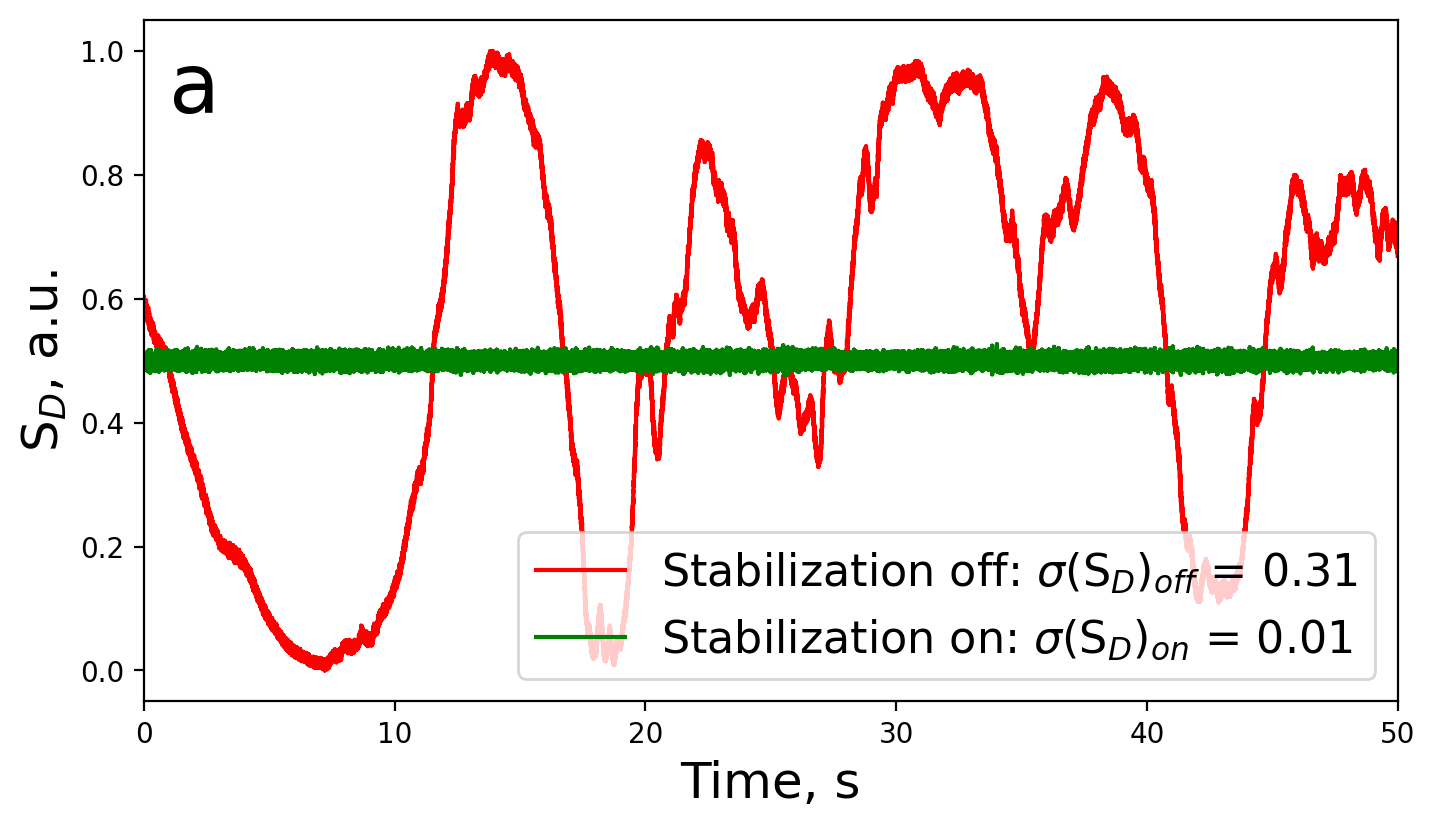}
    \includegraphics[width=0.45\textwidth]{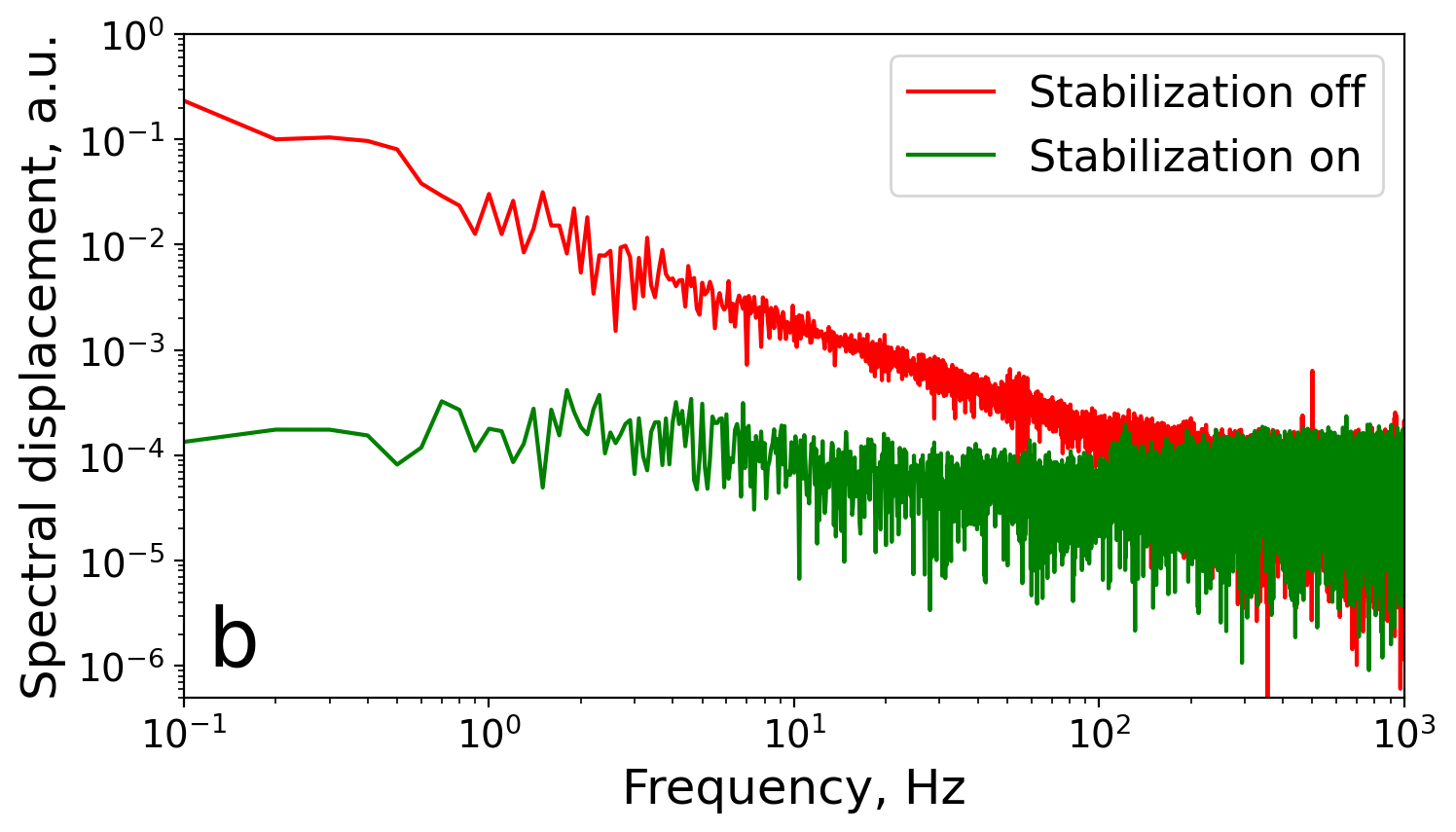}
    \caption{The interference signal S$_D$ normalized by 1 between the W and E leakage channels (a) and its spectra (b) with the relative phase stabilization enabled (solid green) and disabled (solid red).}
    \label{fig:phase_pid}
\end{figure}

The performance of the stabilization system is illustrated by the interference signal S$_D$ normalized by 1 between channels E and W, recorded by photodiode D$^{EW}_{\Delta \phi}$ and shown in Fig.~\ref{fig:phase_pid} for the cases when the relative phase stabilization is disabled (solid red curve) and enabled (solid green curve). Figure \ref{fig:phase_pid} (a) also shows the corresponding values of the S$_D$ RMS for a time of 50 seconds $\sigma$(S$_D$)$_{off}$ and $\sigma$(S$_D$)$_{on}$. The system is effective at frequencies up to 200 Hz (Figure \ref{fig:phase_pid} (b)).

It should be noted that the feedback signal does not account for optical components located downstream of the leakage beam pick-offs. By analogy with automotive engineering, we refer to these components as the ``unsprung mass''. The presence of unsprung mass is unavoidable and inevitably degrades the quality of active phase stabilization at the main focus. In the final XCELS configuration, these elements are planned to be independently stabilized using passive mechanical isolation.



The interference field distribution in the vicinity of the main focus was measured using an original technique based on a fiber-optic probe with a subwavelength aperture, analogous to that of a scanning near-field optical microscope (SNOM) probe \cite{10.1063/1.94865,10.1063/1.336848,Grosjean:10,Bauer2014,photonics10050496}.

\begin{figure}
    \centering
    \includegraphics[width=0.4\textwidth]{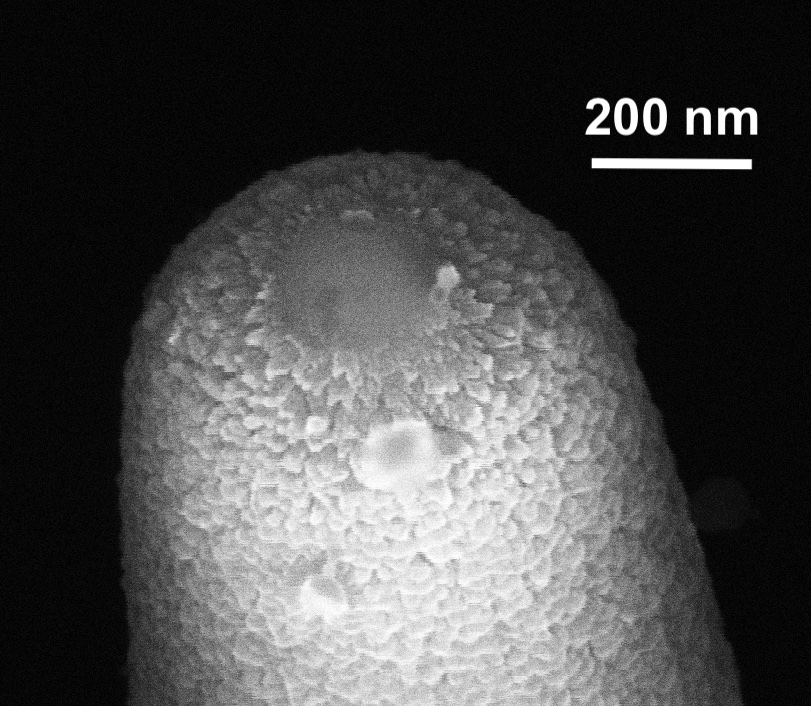}
    \caption{SEM image of the subwavelength probe tip.}
    \label{fig:pin_tip}
\end{figure}

\begin{figure}
    \centering
    \includegraphics[width=0.65\textwidth]{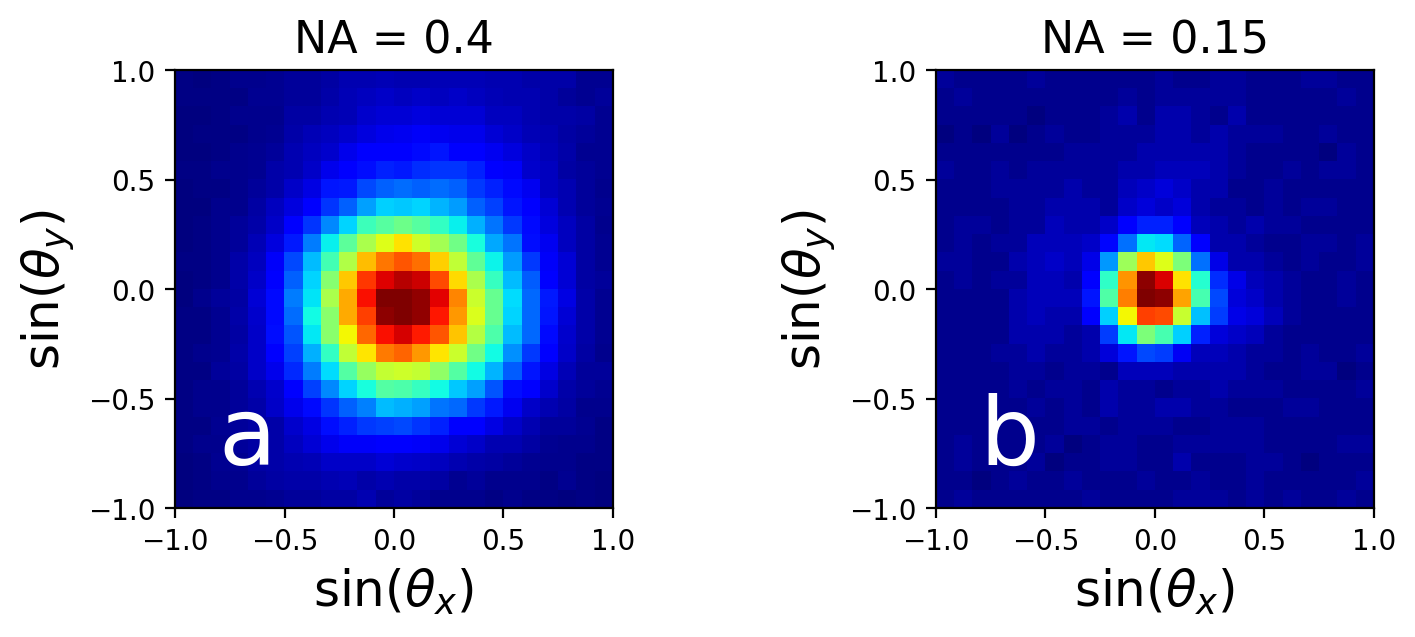}
    \caption{Far-field radiation pattern at the output of the subwavelength probe when illuminated through the FC/PC connector (a), and at the FC/PC connector output when illuminated through the subwavelength aperture (b).}
    \label{fig:zond_diag}
\end{figure}

The probe consisted of a segment of Nufern 780-HP single-mode optical fiber, chemically etched to a sharp tip on one end and coated with an opaque 20-nm-thick metal layer, except for a small subwavelength aperture of approximately 150~nm in diameter at the very apex. An SEM (scanning electron microscope) image of the sharpened probe tip is shown in Fig.~\ref{fig:pin_tip}. The opposite end of the fiber segment was terminated with an FC/PC optical connector. Figure~\ref{fig:zond_diag} shows the far-field intensity distributions of the radiation (a) emitted from the sharpened fiber tip and (b) from the optical connector, when the fiber is illuminated from the opposite end. The sharpened tip exhibits a significantly increased numerical aperture, which indirectly confirms the subwavelength size of the aperture.  

\begin{figure}
    \centering
    \includegraphics[width=0.7\textwidth]{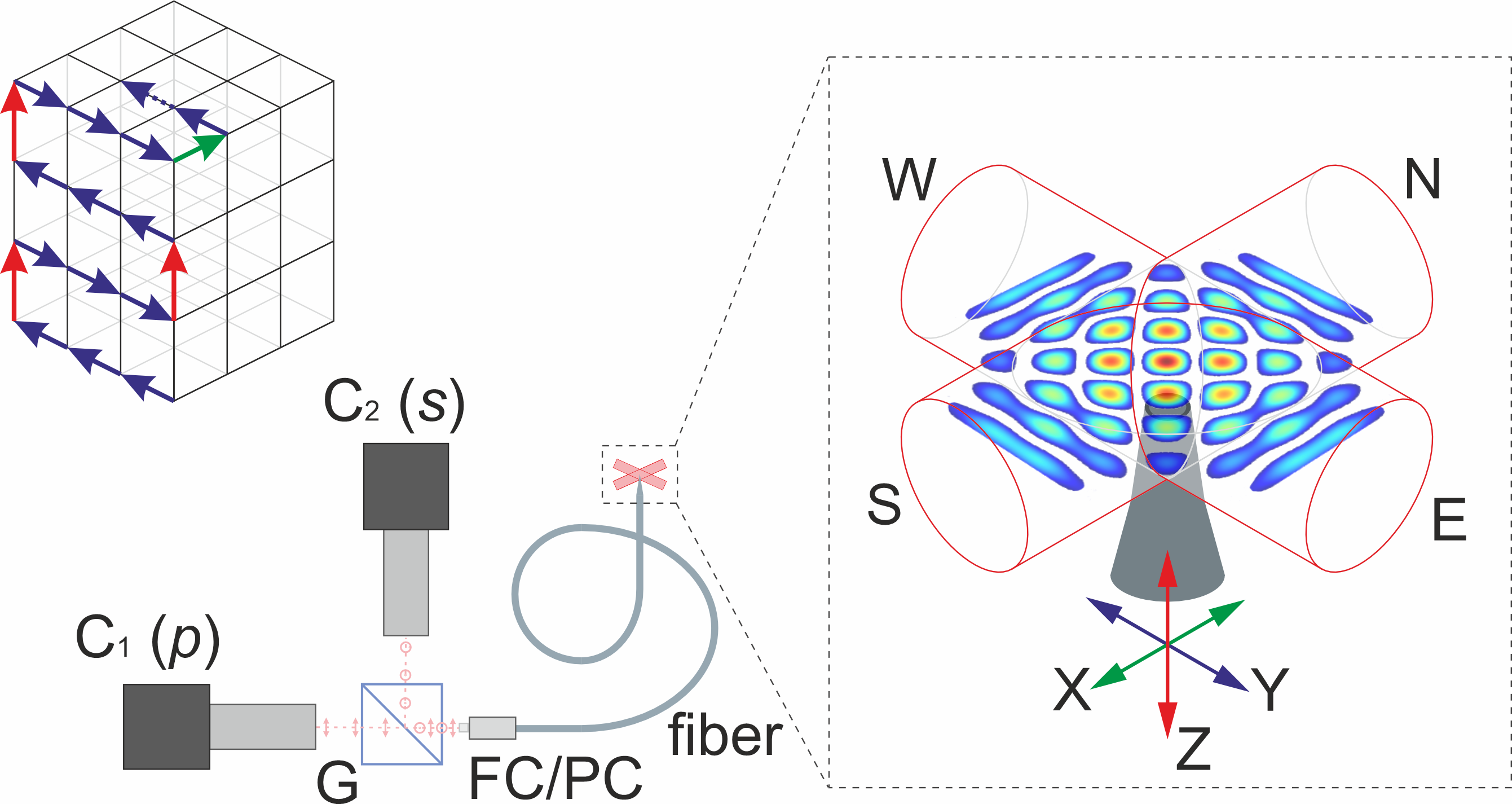}
    \caption{Schematic illustration of the scanning procedure for mapping the optical field distribution in the main focus.}
    \label{fig:zond_scheme}
\end{figure}

Scanning of the region around the main focus with the probe was performed according to the scheme shown in Fig.~\ref{fig:zond_scheme}. The probe was moved step-by-step by a three-axis piezo translation stage along a trajectory similar to that depicted in the upper-left panel.

When radiation falls on the subwavelength aperture of the probe, part of its energy is transformed into the energy of fiber-optic modes traveling freely through the fiber.
Experimentally, it turned out that their amplitude and polarization composition significantly depend on the direction of the wave vector and the polarization of radiation incident on the subwavelength aperture of the probe. In this regard, in order to restore the standing wave configuration, the probe's response to the aggregated effect of each of the channels was preliminarily measured separately. The amplitude and polarization state of the optical signal from the FC/PC fiber connector were recorded using a polarizer G (Glan prism) and two cameras, C$_1$ and C$_2$, equipped with microscope objectives. 
The scanning range was up to 5~\textmu m, with a minimum step size of approximately 50~nm. To minimize vibrations in the scanning system, the probe was repositioned to the next point at a rate not exceeding 5~Hz.


\begin{figure}
    \centering
    \includegraphics[width=0.5\textwidth]{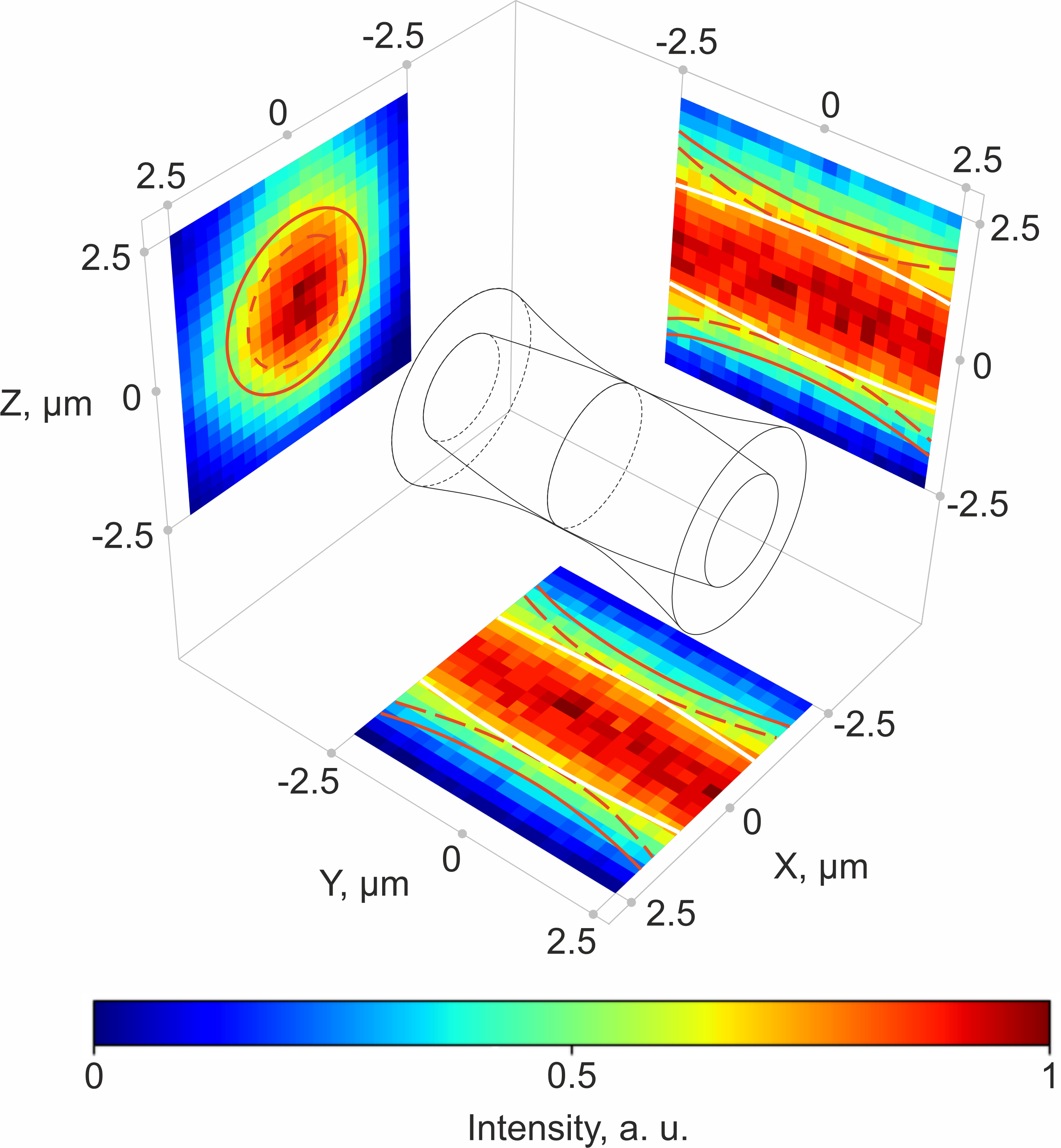}
    \caption{Cross-sections of the E channel waist in the vicinity of MF. Red solid lines are projections of the caustic surface boundaries of the experimentally measured beam with a diameter of 3.5 $\mu$m at the FWHM level. Red dashed lines are projections of the caustic surface boundaries of the beam focused to the diffraction limit by the $f/2$ system (diameter 2.2 $\mu$m at the FWHM level). White lines are projections of the isosurface boundaries at the half-peak intensity level.}
    \label{fig:beam_profile}
\end{figure}

Figure~\ref{fig:beam_profile} shows the beam waist cross-sections of the E channel in the vicinity of the main focus, as measured by the subwavelength probe. The observed broadening of the beam relative to the diffraction-limited size is attributed to wavefront aberrations of the beam. 

\begin{figure}
    \centering
     \includegraphics[width=0.35\textwidth]{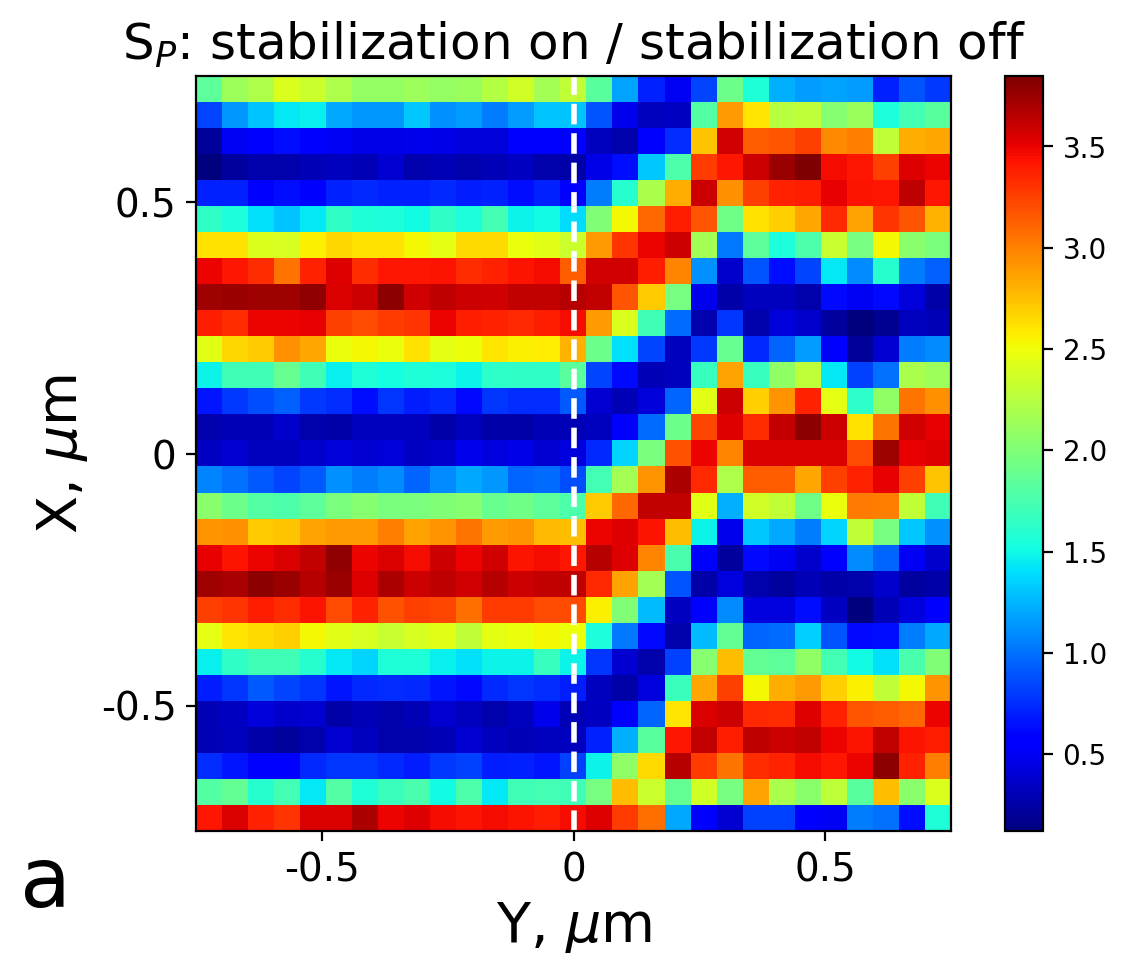}
     \includegraphics[width=0.4\textwidth]{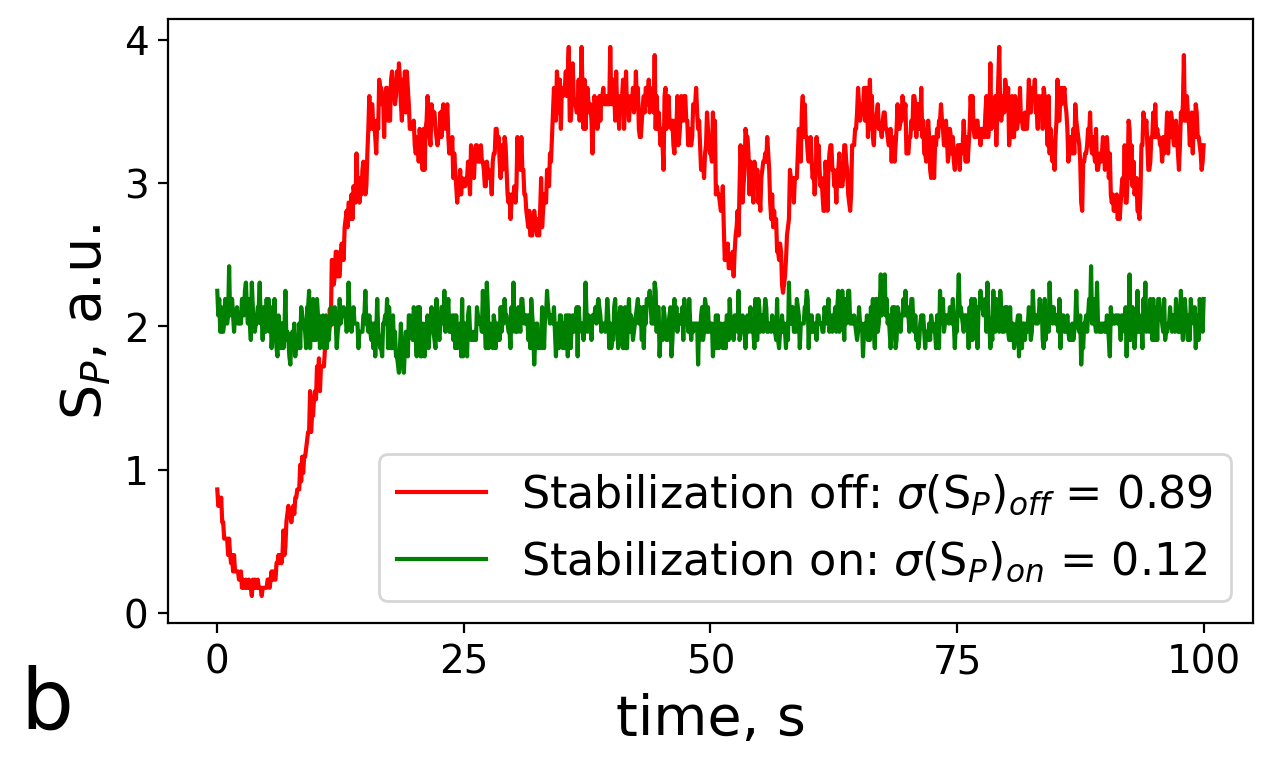}
    \caption{Normalized interference signal S$_P$ during interference of two opposing channels N and S: (a) when scanning with a probe in the XY plane, (b) at a fixed position of the probe.}
    \label{fig:frozen}
\end{figure}

Figure \ref{fig:frozen} demonstrates the stability of the relative phase during the interference of two opposing channels N and S at the main focus. The normalized interference signal S$_P$ recorded using the probe is shown. Figure \ref{fig:frozen} (a) was obtained by scanning with the probe the interference structure in the XY plane, according to the notation introduced in Figure \ref{fig:zond_scheme}. The white vertical dashed line separates the areas of active (left) and inactive (right) stabilization.
The time dynamics of the S$_P$ signal from a stationary probe position is presented in Figure \ref{fig:frozen} (b), for active (green curve) and inactive (red curve) stabilization. The corresponding values of the S$_P$ standard deviation for 100 seconds $\sigma$(S$_P$)$_{on}$ and $\sigma$(S$_P$)$_{off}$ are also given.

\begin{figure}
    \centering
    \includegraphics[width=0.65\textwidth]{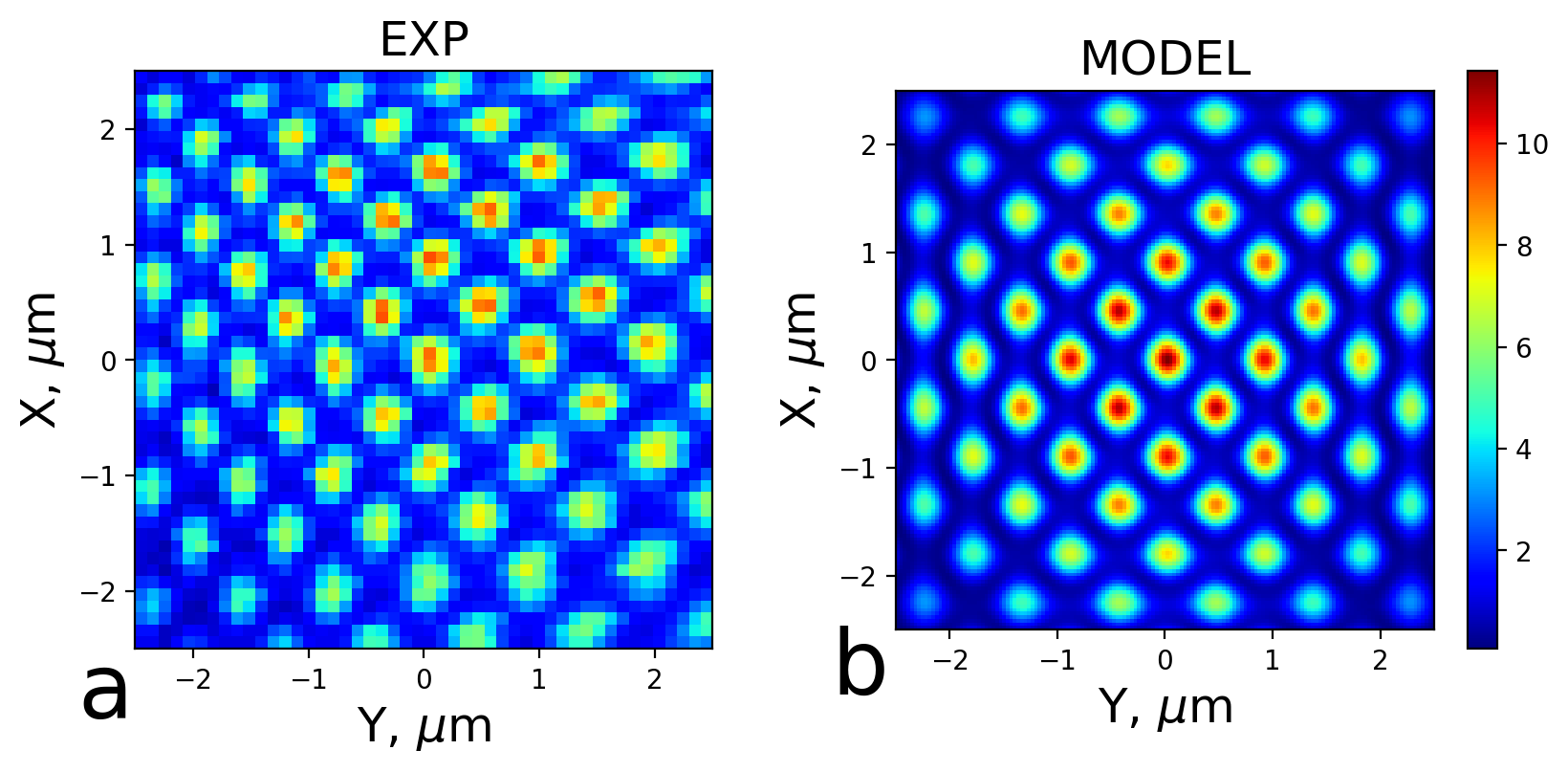}
    \caption{Interference structure of four colliding beams N, S, E and W in the XY plane corresponding to the 4E configuration: experimental measurement using a subwavelength probe (a), result of numerical calculation (b).}
    \label{fig:4beams_field}
\end{figure}

\begin{figure}
    \centering
    \includegraphics[width=0.65\textwidth]{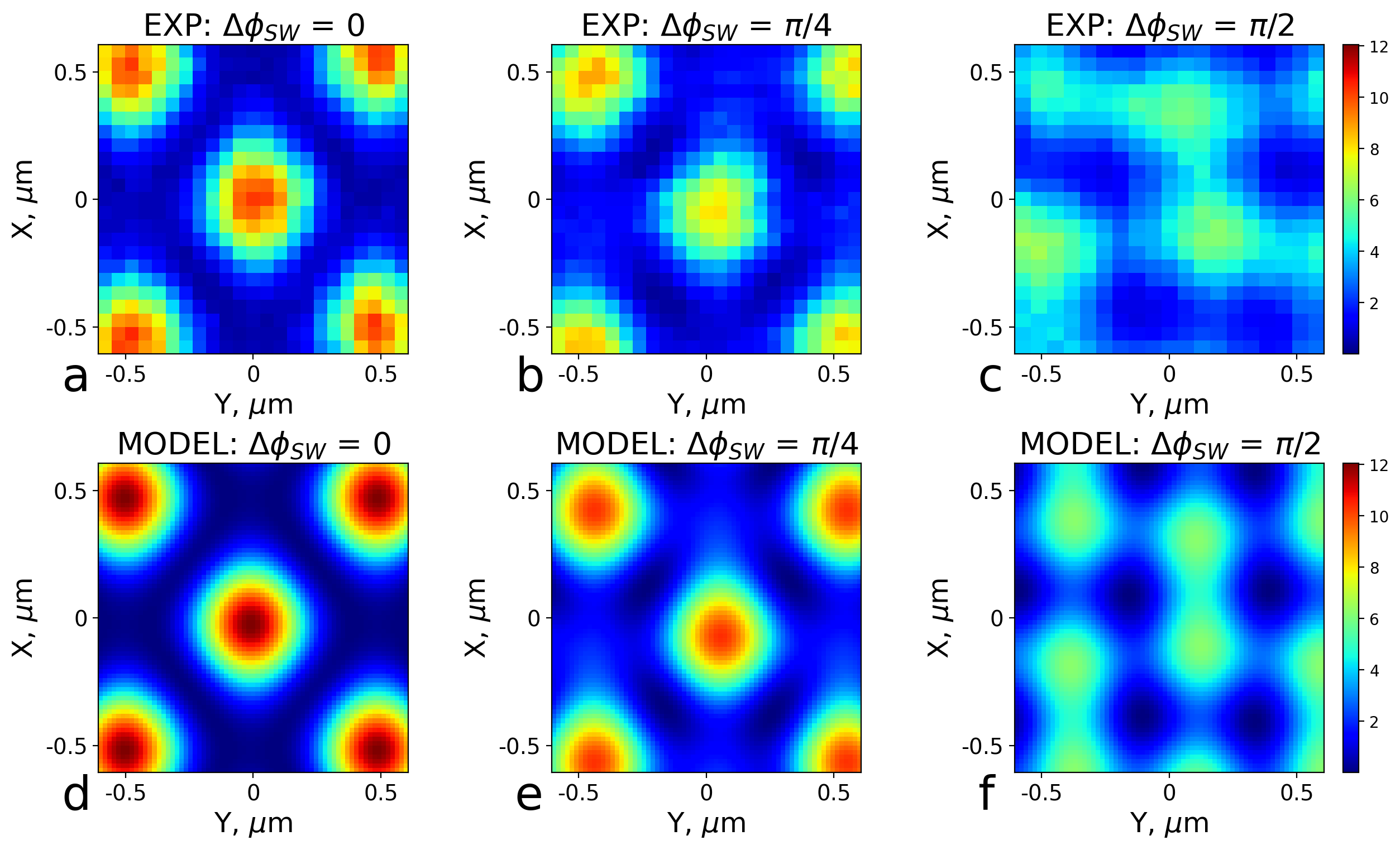}
    \caption{Measured and corresponding calculated interference structures of four counterpropagating beams for: (a,d) zero relative phase between all channels, relative phase shift $\pi/4$ (b,e) and $\pi/2$ (c,f) between pairs of counterpropagating channels.}
    \label{fig:4beams_T_R}
\end{figure}

The field structure when four vertically polarized channels collide, corresponding to the 4E configuration, is presented in Figure \ref{fig:4beams_field}. Figure \ref{fig:4beams_T_R}, along with the 4E-configuration (a and d), shows the structure at the relative phase between opposing pairs of channels $\pi$/4 (b and e) and the 2$\sqrt2$E-configuration (c and f). The numerically calculated field structures in Figures \ref{fig:4beams_field} and \ref{fig:4beams_T_R} take into account the sensitivity of the probe to individual channels and invariants to probe rotation.

Rotating the probe by 90$^\circ$ around the Z-axis led to a redistribution of its response efficiency between individual channels; however, the field distribution structures measured by the probe demonstrated the same response to phase changes between channels, which indicates the invariance of field measurements using a subwavelength probe.


For the XCELS geometry with sharper focusing and a large number of converged channels, one localized antinode will be formed in the vicinity of the MF with an absolute maximum of the squared field, compared to the prototype described in the work, for which several antinodes with almost identical maximum values are observed (see Figures \ref{fig:4beams_field} and \ref{fig:4beams_T_R}). The subwavelength probe will allow to determine the spatial position of the field absolute maximum, which will be provided at zero relative phase of the channels, corresponding to the maximum efficiency of powering the probe.

The aim field structure of the XCELS project is an approximation of a standing electric dipole wave, corresponding to co-directional polarization of all channels and zero relative phase. However, controlling the relative phase and polarization of the channels allows one to approach other configurations, in particular, the magnetic dipole configuration and more complex multipole configurations of standing waves. Disabling/adding individual channels or changing the geometry of their convergence, which violates the equality of the Poynting vector in MF to zero, will allow one to mix a running component into the field.

To quantitatively characterize the scaling from the prototype to the full XCELS project, the following ratios can be identified: the ratio of the squared-field proximity factors ($\xi$) relative to the theoretical dipole limit, which is around 7; the ratio of the total laser channel apertures, approximately 5.5$\times10^{2}$; and the ratio of the combined pulse peak powers, on the order of 10$^{13}$.

Concerning the stabilization systems, previous numerical studies for the XCELS system parameters \cite{bulanov_AO_2025} have shown that high beam pointing stability is essential for achieving efficient coherent combining at the focus. In particular, analysis of various non-idealities (beam aberrations, mirror misalignment, etc.) potentially present in the system indicated that achieving a Strehl ratio of S = 0.9, defined as the ratio of the peak intensity in the focal region in the presence of imperfections to the peak intensity in the ideal case, requires the phase variation amplitude across the beam wavefront not to exceed $\varphi_{0.9} = 0.3$~rad. Furthermore, the relative phase shift between beams must not exceed 0.7~rad to maintain S = 0.9, which corresponds to a path difference of approximately $\lambda/8$.

For the beam pointing stability, this implies that the variation in the angle of incidence on the focusing mirror must not exceed $\alpha_{0.9} = \varphi_{0.9} \cdot \lambda / R$, where $R$ is the beam radius. For the XCELS parameters ($\lambda = 910$~nm, $R = 33$~cm), this yields $\alpha_{0.9} \approx 0.8$~\textmu rad. For the prototype discussed in this work ($\lambda = 910$~nm, $R = 1$~cm), the corresponding limit is $\alpha_{0.9} \approx 27$~\textmu rad. Thus, the achieved angular stability of 0.3--0.8~\textmu rad is well within the requirements for the prototype scale and generally meets the specifications for the full XCELS system.

The phase stabilization system enabled a relative phase stability of better than 0.1~rad at the main focus for the prototype presented herein. This value is an order of magnitude below the 0.7~rad requirement and clearly demonstrates the feasibility of achieving the relative phase stability necessary for efficient coherent beam combining.


In conclusion, the paper presents a four-beam prototype of a dipole focusing system for pulsed femtosecond radiation, developed in support of the XCELS project. 
The setup comprises a system for angular and spatial stabilization of incoming radiation, which is divided into four independent channels equipped with relative phase stabilization subsystems, converged into a common focal point from four distinct directions. Each channel was focused with a numerical aperture of $f/2$, which in total provided a squared electric field amplitude corresponding to $\xi \approx 0.077$ at the main focus.

The interference field structure in the focal region was characterized using an original technique based on scanning with a subwavelength fiber-optic probe. The proposed method demonstrates subwavelength spatial resolution, clearly visualizes standing-wave optical field configurations, and can be effectively used for coherent combining of highly focused beams with a large numerical aperture in a multibeam geometry.

{\bf{ Acknowledgments}}

The experimental part of the work was supported by the Russian Science Foundation under grant no.~25-62-00019. The numerical calculations related to research were carried out within the state assignment of Ministry of Science and Higher Education of the Russian Federation under theme no.~FFUF-2026-0011.

\bibliography{apssamp}

\end{document}